\documentclass[11pt]{article}
\usepackage{amsmath,amsthm,amsfonts,amscd}
\DeclareMathOperator{\Hom}{Hom}
\DeclareMathOperator{\Com}{Com}
\DeclareMathOperator{\CCup}{Cup}
\DeclareMathOperator{\II}{I}
\DeclareMathOperator{\1}{id}
\DeclareMathOperator{\ad}{ad}
\DeclareMathOperator{\dev}{dev}
\DeclareMathOperator{\Ker}{Ker}
\DeclareMathOperator{\IM}{Im}
\newcommand{\NN}{\mathbb{N}}
\newcommand{\RR}{\mathbb{R}}
\newcommand{\CC}{\mathbb{C}}
\newcommand{\EEnd}{\mathcal End}
\newcommand{\CoEEnd}{\mathcal CoEnd}
\newcommand{\EE}{\mathcal E}
\newcommand{\bul}{\bullet}

\renewcommand{\u}{\smile}
\renewcommand{\=}{\doteq}
\renewcommand{\t}{\otimes}

\newcommand{\p}{\partial}
\newtheorem{thm}{Theorem}[section]
\theoremstyle{definition}
 \newtheorem{defn}[thm]{Definition}
\theoremstyle{definition}
 \newtheorem{exam}[thm]{Example}
\begin{document}
\title{\LARGE\bf Invitation to operadic dynamics}
\date{}
\author{\Large Eugen Paal\\ \\
Department of Mathematics, Tallinn University of Technology\\
Ehitajate tee 5, 19086 Tallinn, Estonia\\ \\
E-mail: eugen.paal@ttu.ee}
\maketitle
\thispagestyle{empty}
\begin{abstract}
Based on the Gerstenhaber Theory, clarification is made of how operadic dynamics may be introduced. Operadic observables satisfy the Gerstenhaber algebra identities and their time evolution is governed by operadic 
evolution equation. The notion of an operadic Lax pair is also introduced. As an example, an operadic 
(representation of) harmonic oscillator is proposed.
\end{abstract}

\section{Introduction and outline of the paper}

In 1963, Gerstenhaber invented \cite{Ger} an \emph{operad calculus} 
in the Hochschild complex of an associative algebra;
operads were introduced under the name of \emph{pre-Lie systems}.
In the same year, Stasheff constructed \cite{Sta63} (see also \cite{SnSt})
quite an original geometrical operad, which nowadays is called an \emph{associahedra}.
The notion of an operad was further formalised by May \cite{May72} as a
tool for iterated loop spaces. The main principles of the operad
calculus (brace algebra) were presented by Gerstenhaber and Voronov  \cite{GeVo94,VoGe}. 
Some quite remarkable research activity in the operad theory
and its applications can be observed in the last decade (eg. \cite{Rene,Smi01,MaShSta}). 
It may be said that operads are also becoming an important tool for Quantum Field Theory and 
deformation quantization \cite{Kon}.

Today, much attention is given to static operadic constructions. For dynamical operations one has to prescribe their time evolution. In this paper, based on the Gerstenhaber Theory,  clarification is given on how operadic dynamics may be introduced.

We start from simple algebraic axioms. Basic algebraic constructions
associated with linear operads are introduced. Their properties and the first derivation deviations for the
coboundary operator are presented explicitly. Under certain conditions (a formal associativity constraint), the Gerstenhaber algebra structure appears in the associated cohomology of an operad. 

The operadic dynamics may be introduced by simple and natural analogy with the Hamiltonian version.
Operadic observables satisfy the Gerstenhaber algebra identities  and their time evolution is governed by  
the operadic analogue of the Hamiltonian equations, the  operadic evolution equation. 
The latter describes the time evolution of operations.
In particular, the notion of an operadic Lax pair may be introduced as well.
As an example, an  operadic (representation of) harmonic oscillator is proposed.

\section{Operad}

Let $K$ be a unital associative commutative ring, and let $C^n$
($n\in\NN$) be unital $K$-modules. For \emph{homogeneous} $f\in
C^n$, we refer to $n$ as the \emph{degree} of $f$ and often write
(when it does not cause confusion) $f$ instead of $\deg f$. For
example, $(-1)^f\=(-1)^n$, $C^f\=C^n$ and $\circ_f\=\circ_n$. Also, it
is convenient to use the \emph{reduced} degree $|f|\=n-1$.
Throughout this paper, we assume that $\t\=\t_K$.

\begin{defn}[operad]
A linear (nonsymmetric) \emph{operad} with coefficients in $K$ is a sequence $C\=\{C^n\}_{n\in\NN}$ of unital
$K$-modules (an $\NN$-graded $K$-module), such that the following
conditions are held to be true.
\begin{enumerate}
\item[(1)]
For $0\leq i\leq m-1$ there exist \emph{partial compositions}
\[
  \circ_i\in\Hom(C^m\t C^n,C^{m+n-1}),\qquad |\circ_i|=0
\]
\item[(2)]
For all $h\t f\t g\in C^h\t C^f\t C^g$,
the \emph{composition (associativity) relations} hold,
\[
(h\circ_i f)\circ_j g=
\begin{cases}
    (-1)^{|f||g|} (h\circ_j g)\circ_{i+|g|}f
                       &\text{if $0\leq j\leq i-1$},\\
    h\circ_i(f\circ_{j-i}g)  &\text{if $i\leq j\leq i+|f|$},\\
    (-1)^{|f||g|}(h\circ_{j-|f|}g)\circ_i f
                       &\text{if $i+f\leq j\leq|h|+|f|$}.
\end{cases}
\]
\item[(3)]
Unit $\II\in C^1$ exists such that
\[
\II\circ_0 f=f=f\circ_i \II,\qquad 0\leq i\leq |f|
\]
\end{enumerate}
\end{defn}

In the second item, the \emph{first} and \emph{third} parts of the
defining relations turn out to be equivalent.

\begin{exam}[endomorphism operad {\rm\cite{Ger}}]
\label{HG} Let $L$ be a unital $K$-module and
$\EE_L^n\={\EEnd}_L^n\=\Hom(L^{\t n},L)$. Define the partial compositions
for $f\t g\in\EE_L^f\t\EE_L^g$ as
\[
f\circ_i g\=(-1)^{i|g|}f\circ(\1_L^{\t i}\t g\t\1_L^{\t(|f|-i)}),
         \qquad 0\leq i\leq |f|
\]
Then $\EE_L\=\{\EE_L^n\}_{n\in\NN}$ is an operad
(with the unit $\1_L\in\EE_L^1$) called the \emph{endomorphism operad}
of $L$.

Therefore, algebraic operations can be seen as elements of an endomorphism operad.
\end{exam}

\begin{exam}[coendomorphism operad]
\label{CG}
Let $L$ be a $K$-space and
\[
\overline{\EE}_L^n\={\CoEEnd}_L^n\=\Hom(L,L^{\t n})
\]
Define the partial compositions for $f\t g\in\overline{\EE}_L^f\t\overline{\EE}_L^g$ as
\[
f\circ_i g\=(-1)^{i|g|}(\1_L^{\t i}\t g\t\1_L^{\t(|f|-i)})\circ f,
         \qquad 0\leq i\leq|f|
\]
Then $\overline{\EE}_L\=\{\overline{\EE}_L^n\}_{n\in\NN}$ is an operad (with the unit
$\1_L\in\,\overline{\EE}_L^1$) called the \emph{coendomorphism operad} of $L$.

Therefore, algebraic co-operations can be seen as elements of a coendomorphism operad.
\end{exam}

Just as elements of a vector space are called \emph{vectors}, 
it is natural to call elements of an abstract operad \emph{operations}.

\section{Cup and braces}

Throughout this paper, fix a binary operation $\mu\in C^2$ in an operad $C$.

\begin{defn}
The \emph{cup-multiplication} $\u\:C^f\t C^g\to C^{f+g}$ is defined
by
\[
f\u g\=(-1)^f(\mu\circ_0 f)\circ_f g\in C^{f+g},
\qquad|\smile|=1
\]
The pair $\CCup C\=\{C,\u\}$ is called a $\u$-algebra (cup-algebra) of $C$.
\end{defn}

\begin{exam}
For the endomorphism operad (Example \ref{HG}) $\EE_L$ one has
\[
f\u g=(-1)^{fg}\mu\circ(f\t g),
      \qquad \mu\t f\t g\in\EE_L^2\t\EE_L^f\t\EE_L^g
\]
\end{exam}

\begin{defn}
The \emph{total composition} $\bul\:C^f\t C^g\to C^{f+|g|}$ is defined by
\[
f\bul g\=\sum_{i=0}^{|f|}f\circ_i g\in C^{f+|g|},
\qquad |\bul|=0
\]
The pair $\Com C\=\{C,\bul\}$ is called the \emph{composition algebra} of $C$.
\end{defn}

\begin{defn}[tribraces]
Define the Gerstenhaber \emph{tribraces} $\{\cdot,\cdot,\cdot\}$
as a double sum
\[
\{h,f,g\}\=\sum_{i=0}^{|h|-1}\sum_{i+f}^{|f|+|h|}(h\circ_i f)\circ_j g\in C^{h+|f|+|g|},
\quad|\{\cdot,\cdot,\cdot\}|=0
\]
\end{defn}
\begin{defn}[tetrabraces]
The \emph{tetrabraces} $\{\cdot,\cdot,\cdot,\cdot\}$ are defined by
\[
\{h,f,g,b\}\=\sum_{i=0}^{|h|-2}\sum_{j=i+f}^{|h|+|f|-1}\sum_{k=j+g}^{|h|+|f|+|g|}
((h\circ_{i}f)\circ_{j}g)\circ_{k}b\in C^{h+|f|+|g|+|b|},
\quad |\{\cdot,\cdot,\cdot,\cdot\}|=0
\]
\end{defn}
\noindent
It turns out that
\[
f\u g=(-1)^f\{\mu,f,g\}
\]
In general, $\CCup C$ is a
\emph{non-associative} algebra. By denoting $\mu^{2}\=\mu\bul\mu$, it turns
out that the associator in $\CCup C$ reads
\[
(f\smile g)\smile h-f\smile(g\smile h)=\{\mu^{2},f,g,h\}
\]
Therefore the \emph{formal associator} (\emph{micro-associator}) $\mu^{2}$ is an
obstruction to the associativity of $\CCup C$. For an endomorphism operad
$\EE_L$, the ternary operation $\mu^{2}$ also reads as an associator:
\[
\mu^{2}=\mu\circ(\mu\t\1_L-\1_L\t\mu),\qquad\mu\in\EE_L^2
\]

\section{Associated graded Lie algebra}

In an operad $C$, the Getzler identity
\[
(h,f,g) \=(h\bul f)\bul g-h\bul(f\bul g)
         =\{h,f,g\}+(-1)^{|f||g|}\{h,g,f\}
\]
holds, which easily implies the Gerstenhaber identity
\[
(h,f,g)=(-1)^{|f||g|}(h,g,f)
\]
The  \emph{Gerstenhaber brackets} $[\cdot,\cdot]$ are defined in $\Com C$ as a graded commutator by
\[
[f,g]\=f\bul g-(-1)^{|f||g|}g\bul f=-(-1)^{|f||g|}[g,f],\qquad|[\cdot,\cdot]|=0 \tag{G1}
\]
The \emph{commutator algebra} of $\Com C$ is denoted as
$\Com^{-}\!C\=\{C,[\cdot,\cdot]\}$. By using the Gerstenhaber identity, one
can prove that $\Com^-\!C$ is a \emph{graded Lie algebra}. The Jacobi
identity reads
\[
(-1)^{|f||h|}[[f,g],h]+(-1)^{|g||f|}[[g,h],f]+(-1)^{|h||g|}[[h,f],g]=0 \tag{G2}
\]

\section{Coboundary operator}

In an operad $C$, by using the Gerstenhaber brackets,   
a \emph{(pre-)coboundary} operator $\p\=\p_\mu$ may be defined by
\begin{align*}
\p f&\=\ad_\mu^{right}f\=[f,\mu]\=f\bul\mu-(-1)^{|f|}\mu\bul f \\
      &\,\,=f\u\II+f\bul\mu+(-1)^{|f|}\,\II\u f,\qquad \deg\p=+1=|\p|
\end{align*}
It follows from the Jacobi
identity in $\Com^{-}\!C$ that $\p$ is a (right) derivation of
$\Com^{-}\!C$,
\[
\p[f,g]=(-1)^{|g|}[\p f,g]+[f,\p g]
\]
and one has the commutation relation
\[
[\p_f,\p_g]\=\p_f\p_g-(-1)^{|f||g|}\p_g\p_f=\p_{[g,f]}
\]
Therefore, since $|\mu|=+1$ is \emph{odd}, then
\[
\p_\mu^{2}=\frac{1}{2}[\p_\mu,\p_\mu]=
\frac{1}{2}\p_{[\mu,\mu]}=\p_{\mu\bul\mu}=\p_{\mu^{2}}
\]
Here we assumed that $2\neq0$, the proof for an arbitrary characteristic may be found from 
\cite{KPS}. But $\p$ need not be a derivation of $\CCup C$, and $\mu^{2}$ again appears
as an obstruction:
\[
\p(f\smile g)-f\smile\p g-(-1)^{g}\p f\smile g
=(-1)^{g}\{\mu^{2},f,g\}
\]

\section{Derivation deviations}

The \emph{derivation deviation} of $\p$ over $\bul$ is defined by
\[
(\dev_\bul\p)(f\t g)
   \=\p(f\bul g)-f\bul\p g-(-1)^{|g|}\p f\bul g
\]

\begin{thm}
\label{first}
In a pre-operad $C$, one has
\[
(-1)^{g}(\dev_\bul\p)(f\t g)=f\u g-(-1)^{fg}g\u f
\]
\end{thm}
\begin{proof}
The full proof is presented in \cite{KP}.
\end{proof}
\noindent
The derivation deviation of $\p$ over $\{\cdot,\cdot,\cdot\}$ is defined by
\begin{align*}
(\dev_{\{\cdot,\cdot,\cdot\}}\p)(h\t f\t g)
    \=\p\{h,f,g\}
     &-\{h,f,\p g\}\\
     &-(-1)^{|g|}\{h,\p f,g\}-(-1)^{|g|+|f|}\{\p h,f,g\}
\end{align*}

\begin{thm}
\label{second}
In a pre-operad $C$, one has
\[
(-1)^{g}(\dev_{\{\cdot,\cdot,\cdot\}}\p)(h\t f\t g)=
    (h\bul f)\u g+(-1)^{|h|f}f\u(h\bul g)-h\bul(f\u g)
\]
\end{thm}
\begin{proof}
The full proof is presented in \cite{KPS}.
\end{proof}
Therefore the \emph{left} translations in $\Com C$ are not
derivations of $\CCup C$, the corresponding deviations are related to
$\dev_{\{\cdot,\cdot,\cdot\}}\p$. It turns out that the \emph{right}
translations in $\Com C$ are derivations of $\CCup C$,
\[
(f\u g)\bul h=f\u(g\bul h)+(-1)^{|h|g}(f\bul h)\u g
\]
By combining this formula with the one from Theorem \ref{second} we obtain

\begin{thm}
\label{second*}
In a pre-operad $C$, one has
\[
(-1)^{g}(\dev_{\{\cdot,\cdot,\cdot\}}\p)(h\t f\t g)=
    [h,f]\u g+(-1)^{|h|f}f\u[h,g]-[h,f\u g]
\]
\end{thm}

\section{Gerstenhaber Theory}

Now, clarification can be supplied to show how the Gerstenhaber algebra can be associated with a
linear operad. If (formal associativity) $\mu^{2}=0$ holds, then
$\p^{2}=0$, which in turn implies $\IM\p\subseteq\Ker\p$. Then one can
form an associated cohomology ($\NN$-graded module) $H(C)\=\Ker\p/\IM\p$ with
homogeneous components
\[
H^{n}(C)\=\Ker(C^{n}\stackrel{\p}{\rightarrow}C^{n+1})/
           \IM(C^{n-1}\stackrel{\p}{\rightarrow}C^{n})
\]
where, by convention, $\IM(C^{-1}\stackrel{\p}{\rightarrow}C^{0})\=0$.
Also, in this ($\mu^{2}=0$) case, $\CCup C$ is \emph{associative},
\[
(f\u g)\u h=f\u(g\u h) \tag{G3}
\]
and $\p$ is a \emph{derivation} of $\CCup C$. Remember from previously that
$\Com^{-}\!C$ is a graded Lie algebra and $\p$ is a derivation of
$\Com^{-}\!C$. Due to the derivation properties of $\p$, the
multiplications $[\cdot,\cdot]$ and $\smile$ induce corresponding (factor)
multiplications on $H(C)$, which we denote by the same symbols. Then
$\{H(C),[\cdot,\cdot]\}$ is a \emph{graded Lie algebra}. It follows from
Theorem \ref{first} that the induced $\smile$-multiplication on $H(C)$ is
\emph{graded commutative},
\[
f\smile g=(-1)^{fg}g\smile f \tag{G4}
\]
for all $f\t g\in H^{f}(C)\t H^{g}(C)$, hence $\{H(C),\smile\}$ is an
\emph{associative graded commutative} algebra.
It follows from Theorem \ref{second*} that the  \emph{graded Leibniz rule}
holds,
\[
[h,f\u g]=[h,f]\u g+(-1)^{|h|f}f\u[h,g] \tag{G5}
\]
for all $h\t f\t g\in H^{h}(C)\t H^{f}(C)\t H^{g}(C)$. At last, it is also
relevant to note that
\[
0=|[\cdot,\cdot]|\neq|\smile|=1  \tag{G6}
\]
In this way, the triple $\{H(C),\smile,[\cdot,\cdot]\}$ turns out
to be a \emph{Gerstenhaber algebra} \cite{GGS92Am}.
The defining relations of a Gerstenhaber algebra are (G1)-(G6).

In the case of an endomorphism operad, the Gerstenhaber algebra
structure appears on the Hochschild cohomology of an associative algebra
\cite{Ger}. This is the essence of the Gerstenhaber Theory.

In particular, in the case of a coendomorphism operad, the Gerstenhaber algebra
structure appears on the Cartier cohomology of a coassociative coalgebra.

\section{Operadic dynamics}

Assume that $K\doteq\RR$ or $K\doteq\CC$.
It is known that the Poisson algebras can be seen as an algebraic abstraction of mechanics.
Consider the following figurative commutative diagram:
\[
\begin{CD}
\text{Poisson algebras}@<\text{algebra}<<\text{mechanics}\\
      \wr                       @.                        \wr\\
\text{Gerstenhaber algebras}@<\text{algebra}<<\text{operadic mechanics}
\end{CD}
\]
Concisely speaking, \emph{operadic observables} are elements of a  Gerstenhaber algebra.
The time evolution of an operadic operadic obervable $f$ is governed by the  
\emph{operadic evolution equation}
\[
\dfrac{df}{dt}=[H,f]\=H\bul f-(-1)^{|H||f|}f\bul H
\]
with the (model-dependent)  \emph{operadic Hamiltonian} $H$. 
The most simple assumption for its degree is
\begin{gather*}
\left|\dfrac{d}{dt}\right|=|H|=0
\quad
\Longrightarrow
\quad
[H,f]\=H\bul f-f\bul H
\end{gather*}
In particular, 
\[
|H|=|f|=0
\quad
\Longrightarrow
\quad
[H,f]=H\circ f-f\circ H
\]
and in this case one finds the well-known evolution equation
\[
\dfrac{df}{dt}=[H,f]\=H\circ f-f\circ H
\]

In this way one can describe the time evolution of operations.
In particular, one can propose the

\begin{defn}[operadic Lax pair]
Allow a classical dynamical system to be described by the evolution equations 
\[
\dfrac{dx_i}{dt}=f_i(x_1,\dots,x_n),\quad i=1,\dots,n
\]
An \emph{operadic Lax pair} is a pair of homogeneous operations $L,M\in C$, 
such that the above system of evolution equations is equivalent to the
\emph{operadic Lax equation}
\[
\dfrac{dL}{dt}=[M,L]\=M\bul L-(-1)^{|M||L|}L\bul M
\]
Evidently, $|M|=0$ is the most simple assumption and the degree constraints $|M|=|L|=0$
give rise to ordinary Lax pair \cite{Lax68}.
\end{defn}

Endomorphism and co-endomorphism operads are the most natural objects for modelling 
operadic dynamical systems.

Surprisingly, examples are at hand. By using the Lax pairs one may extend these 
to operadic area via the operadic Lax equation. 

\begin{exam}[operadic harmonic oscillator]
Consider the classical Lax pair for the harmonic oscillator:
\[
L_{cl}=\begin{pmatrix}
p&\omega q\\
\omega q &-p
\end{pmatrix},
\qquad
M=
\begin{pmatrix}
0&-\omega/2\\
\omega/2 &0
\end{pmatrix}
\]
Since the Hamiltonian is
\[
H(q,p)=\frac{1}{2}(p^2+\omega^2q^2)
\]
one can use the Hamiltonian canonical equations 
\[
\dfrac{dq}{dt}=\dfrac{\partial H}{\partial p}=p,
\quad
\dfrac{dp}{dt}=-\dfrac{\partial H}{\partial q}=-\omega^2q
\]
to obtain
\[
\dfrac{dL}{dt}
=\dfrac{\partial L}{\partial q}\dfrac{dq}{dt}+\dfrac{\partial L}{\partial p}\dfrac{dp}{dt}
=p\dfrac{\partial L}{\partial q}-\omega^2q\dfrac{\partial L}{\partial p}
\]
Therefore the linear partial differential equation for the operadic variable $L(q,p)$ reads
\[
p\dfrac{\partial L}{\partial q}-\omega^2q\dfrac{\partial L}{\partial p}=M\bul L- L\bul M
\]
By integrating one gains a sequence of operations  called an \emph{operadic (representation of) harmonic oscillator}.
\end{exam}

\section*{Acknowledgement}
Research was in part supported by the Estonian Science Foundation, Grants 5634 and 6912.
The author is grateful to Piret Kuusk for reading the preliminary manuscript, 
and for contributing valuable remarks and discussions.

\end{document}